\documentclass{aa}
\def\puncspace{\ifmmode\,\else{\ifcat.\C{\if.\C\else\if,\C\else\if?\C\else%
\if:\C\else\if;\C\else\if-\C\else\if)\C\else\if/\C\else\if]\C\else\if'\C%
\else\space\fi\fi\fi\fi\fi\fi\fi\fi\fi\fi}%
\else\if\empty\C\else\if\space\C\else\space\fi\fi\fi}\fi}
\def\SP{\let\\=\empty\futurelet\C\puncspace}
\def\etal{{\it et al.\/}\ }

\def\eg{{\it e.g.\/}\rm\ }
\def\lsim{~\rlap{$<$}{\lower 1.0ex\hbox{$\sim$}}}
\def\gsim{~\rlap{$>$}{\lower 1.0ex\hbox{$\sim$}}}
\def\void#1{{}}

\usepackage{graphicx}

\def\etal{{\it et al.\/}\ }

\usepackage{longtable,threeparttable}
\begin{document}
\title{ESO Imaging Survey}  {\subtitle{Pre-FLAMES Survey: 
Observations of Selected Stellar Fields\thanks{Based on observations
collected at the European Southern Observatory, La Silla, Chile within
program ESO 164.O-O561.}.}

\author{ Y. Momany       \inst{1,2}
 \and B. Vandame         \inst{1} 
 \and S. Zaggia          \inst{1,3}
 \and R. P. Mignani      \inst{1}
 \and L. da Costa        \inst{1}
 \and S. Arnouts        \inst{1}
 \and M.A.T. Groenewegen \inst{1}
 \and E. Hatziminaoglou  \inst{1}
 \and R. Madejsky        \inst{1,5}
 \and C. Rit\'e          \inst{1,6}
 \and M. Schirmer        \inst{4}
 \and R. Slijkhuis       \inst{1}
}
\offprints{Yazan Momany, e-mail: ymomany@eso.org}
\institute{
     European Southern Observatory, Karl-Schwarzschild-Str. 2, 
     D-85748 Garching b. M\"unchen, Germany
\and Dipartimento di Astronomia, Universit\'a di Padova,
     Vicolo dell'Osservatorio 5, I-35122 Padova, Italy 
\and Osservatorio Astronomico di Trieste, Via G. B. Tiepolo 11,
     I-34131 Trieste, Italy
\and Max-Planck Institut f\"ur Astrophysik, Karl-Schwarzschild-Str. 1,
     D-85748 Garching b. M\"unchen, Germany
\and Universidade Estadual de Feira de Santana, Campus
     Universit\'ario, Feira de Santana, BA, Brazil
\and Observat\'orio Nacional, Rua Gen. Jos\'e Cristino 77, 
     Rio de Janerio, R.J., Brasil }
	
\date{Received July 25, 2001; accepted ......, 2001}

\abstract{This paper presents the first set of fully calibrated images
and associated stellar catalogs of the Pre-FLAMES survey being carried
out by the ESO Imaging Survey (EIS) project.  The primary goal of this
survey is  to provide  the ESO  community  with data  sets  from which
suitable target lists can be extracted for follow-up observations with
the new VLT  facility  FLAMES    (Fiber  Large Array Multi     Element
Spectrograph). For this purpose 160  stellar fields have been selected
for  observations in $B$,  $V$ and  $I$  using the $8$k$\times8$k Wide
Field Imager (WFI) at the MPG/ESO 2.2 m  telescope at La Silla. Out of
those over 100 fields have already been observed. The list of selected
fields includes  open  clusters,  globular clusters,  regions  in  the
Galaxy  bulge, dwarf spheroidal galaxies in  the vicinity of the Milky
Way,  contiguous regions of SMC  and  LMC and  few nearby clusters  of
galaxies. The present paper discusses the results obtained for a small
subset of these data which include four open clusters (M~67, NGC~2477,
NGC~2506 and Berkeley~20) and two regions of the SMC.  These data have
been used  to assess the  observing strategy adopted, a combination of
short- and long-exposures, and to define suitable reduction techniques
and procedures for  the preparation of  input catalogs for FLAMES.  In
order to minimize light losses  due to misplacements of FLAMES fibers,
the  astrometric calibration of crowded  stellar  fields is a critical
issue.  The impact  of  different  swarping techniques  and  different
reference   catalogs on the astrometric   calibration of the images is
evaluated  and   compared  to those of     other  authors.  From  this
comparison one finds that both USNO~2.0  and recently released GSC~2.2
yield comparable results with  the positional differences having a rms
of  about $0.15$~arcsec well  within  the requirements ($0.2$  arcsec)
specified by  the FLAMES science team.  The  internal  accuracy of the
astrometry is estimated  to be $\lsim$~0.1~arcsec primarily limited by
the reference catalog  used.    The  major difference  between   these
catalogs is the systematic variation of the  positional residuals as a
function  of the apparent magnitude of  the objects,  with the GSC~2.2
yielding by  far the best  results. The astrometric calibration of the
images presented here is based on the USNO~2.0 catalog because not all
fields considered  are covered by  the current release of the GSC~2.2.
Future EIS calibrations  will be done using the  GSC~2.2 catalog.  The
extraction and photometric measurements of stellar sources are carried
out using a  PSF fitting technique.  Comparison with results available
in literature shows  that  the photometric  measurements are in   good
agreement, apart from possible  zero-point offsets, with the magnitude
differences having  a scatter of  $\sim 0.06$~mag at  $V=20$~mag. This
demonstrates that the data  allow for the  selection of robust targets
down to the expected spectroscopic limit of FLAMES. The combination of
catalogs extracted  from  the short  and long-exposures allows  one to
produce color-magnitude  diagrams  (CMD) spanning $\sim13$~mag in  $V$
and reaching a limiting   magnitude of $V\sim22-23$.  These  data have
also been combined   with data from   the Two Micron All Sky Survey
(2MASS) survey allowing  for a
better color-based,  object classification and  target selection.  The
Pre-Flames   (PF)  survey data meet the   requirements  of FLAMES, and
provide a good starting  point for  detailed  studies of the  examined
systems.  
The images and catalogs  presented here are publicly  available and can be
requested from the URL address ``http://www.eso.org/eis''.
\keywords{catalogs-surveys-stars-FLAMES} }

\maketitle

\section{Introduction}

The ESO Imaging Survey (EIS) project was conceived with the purpose of
producing data  sets  matching  the   foreseen scientific  goals   and
requirements  of different VLT instruments  (\eg Renzini  and da Costa
1997). 
With
this  in mind, EIS  has been for the past   two years carrying out the
Deep    Public  Survey (DPS),  an  optical/infrared    deep survey  of
high-galactic latitude  fields and the  so-called Pre-Flames (PF). The
latter survey   consists of selected stellar   fields well  matched to
VIMOS and  FLAMES   (Fibre Large  Array   Multi Element  Spectrograph,
Pasquini \etal\ 2000) capabilities, respectively.

FLAMES, which will be installed on the A Nasmyth platform of the VLT
Kueyen telescope, consists of a fiber positioner, a dedicated
fiber-fed spectrograph (GIRAFFE) and a fiber link to the UVES
spectrograph located on the B Nasmyth platform. The fiber positioner
covers a corrected field of view of $\simeq25$~arcmin in diameter.
GIRAFFE will be fed by fibers in one of the following ways: (a) 130
fibers, 1.2~arcsec in diameter, in the MEDUSA configuration; (b) $15$
deployable integral field units (IFUs, $2\times3$ square arcsec with
20 fibers each); (c) one central large unit (ARGUS, $11.5\times7.3$
square~arcsec with 308 fibers).  GIRAFFE will have two gratings with a
resolving power of $R=5000-9000$ and $R=15000-25000$, depending on the
fiber mode used. The corresponding spectral coverage will be
$\sim50-120$~nm and $\sim20-50$~nm, respectively. The performance of
the instrument can be illustrated by the following example: an
exposure of one-hour with GIRAFFE in the Medusa mode of a $V=20$~mag
point source at $R=15000$ is expected to yield a spectrum with a
signal-to-noise ratio of 15 per resolution element. UVES will be fed
by 8 fibers and will only use the red arm with a resolving power of
$R\sim45000$ and a spectral coverage of $200$~nm.  An important
feature of the FLAMES setup is that it will allow for simultaneous
observations with GIRAFFE and UVES.

The relatively small diameter of the fibers and the lack of an imaging
mode require the preparation of target lists with accurate astrometry.
The positional errors have to be small ($\lsim0.2$ arcsec) in order to
minimize the off-center losses.  It is estimated that misplacing a fiber
by $\sim0.5$ arcsec, during observations under seeing conditions typical
of Paranal ($\sim0.7$~arcsec) would decrease the collected flux by
$\simeq50\%$.  In addition, to take full advantage of GIRAFFE, 
multi-color source catalogs with reliable photometry (\eg $\sim0.03$~mag
at $V=20$) over the large field-of-view of FLAMES are required for an
adequate pre-selection of spectroscopic targets and their subsequent
analysis.

Foreseeing the need for building suitable data sets for FLAMES, ESO's
Working Group for public surveys recommended the EIS project to carry
out an imaging survey of selected dense stellar fields, the so-called
Pre-Flames(PF) Survey.  The survey is being conducted with the
wide-field imager (WFI) at the MPG/ESO $2.2$~m telescope, with a field
of view ($34\times33$~arcmin) comparable to that of FLAMES
($\simeq25$~arcmin in diameter). Like in the case of other public
surveys carried out by EIS the ultimate goal has been not only to
gather imaging data, but develop and test procedures to produce
science grade products in the form of fully calibrated images and
multi-color stellar catalogs, from which samples for observations with
FLAMES can be extracted. The survey was designed to observe a suitable
number of fields required for commissioning and first year of
operation of FLAMES.  The fields should have surface densities $>1000$
objects per square degree at the magnitude limit of FLAMES. Such
fields will provide enough targets for the 130 fibers available in the
Medusa mode.  Considering that in a typical night the MEDUSA mode
can produce around $1000$ stellar spectra in five to ten different
fields (Pasquini 2000) this implies that approximately $500$ stellar
fields per year can be observed with FLAMES.  To this end a total of
160 fields were selected for observations from suggestions of end
users as compiled by the FLAMES team. Test runs were conducted during
the first semester of $1999$, as part of the EIS Pilot Survey. These
earlier data helped to defined the observing strategy of the PF survey
which started in October 1999.

This paper presents the first set of data produced, which is used to
assess the procedures being adopted in carrying out the observations
and data reduction, to evaluate the results and to illustrate the
scientific potential of the data once combined with the spectroscopic
observations using FLAMES.  This paper is organized as follows:
Section~\ref{S_observation} describes the observations conducted in
the period November $27-29$, $2000$, one of 10 PF runs
conducted so far. Section~\ref{S_process} describes the data reduction
steps, while Section~\ref{S_astrometry} presents a detailed discussion
of the astrometric calibration, including the results of tests
conducted using different swarping techniques and reference
catalogs. In this paper only the $B$ and $V$ images are presented,
leaving to a subsequent paper the description of the reduction of $I$
band which requires a special treatment because of the strong and
variable effects of fringing. The methods used in the preparation of
the final photometric catalog which includes source extraction in
crowded fields, magnitude measurements based on PSF fitting, merging
of catalogs extracted from short- and long-exposures and from
different passbands are discussed in Section~\ref{S_photo}.
Section~\ref{S_release} describes the science products being released
which can be requested from the URL ``http://www.eso.org/eis''. A
preliminary evaluation of the data is presented in
Section~\ref{S_discusion} which presents and discusses the
color-magnitude diagrams for the different stellar systems. Finally,
in Section~\ref{S_summary} the main results of the present paper are
summarized and the future implementations are outlined.

\begin{table*}[t]
\begin{center}
\caption{Log of the  November $27-29$, $2000$ observing run.}
\begin{tabular}{lccccccc} \hline\hline
&&&&&&\\[-9pt]
 Target   &      RA        &        Dec      & Night  & Filter & Exp.Time & seeing \\
 EIS name &(h\,\,\, m\,\,\, s)&(d\,\,\, m\,\,\, s)&   &        & (sec)    &(arcsec)\\[3pt]
\hline
&&&&&&\\[-9pt]
 OC3&05\, 32\, 57.9& $+$00\, 13\, 04 & 27 Nov & $B$ & $2\times240 + 1\times30$ & $1.09-0.99$ \\ 
 (Berkeley~20)     &              &                 &        & $V$ & $2\times240 + 1\times30$ & $1.02-1.14$ \\[6pt]
 SMC~5  & 00\, 56\, 45.0 & $-$72\, 19\, 00 & 28 Nov & $B$ & $2\times240 + 1\times30$ & $1.36-1.29$ \\ 
 (SMC)  &                &                 &        & $V$ & $2\times240 + 1\times30$ & $1.33-1.17$ \\[6pt]
 OC14 & 08\, 00\, 10.7 & $-$10\, 47\, 17 & 28 Nov & $B$ & $2\times240 + 1\times30$ & $0.90-0.79$ \\ 
 (NGC~2506)   &                &                 &        & $V$ & $2\times240 + 1\times30$ & $0.95-1.02$ \\[6pt]
 SMC~6     & 01\, 03\, 35.0 & $-$72\, 19\, 00 & 29 Nov & $B$ &
$2\times240 + 1\times30$ & $1.29-1.22$ \\ 
 (SMC)  &                &                 &        & $V$ & $1\times240 + 1\times30$ & $1.29-1.33$ \\[6pt]
 OC12 & 07\, 52\, 16.7 & $-$38\, 32\, 48 & 29 Nov & $B$ & $2\times240 + 1\times30$ & $1.23-1.09$ \\ 
 (NGC~2477)   &                &                 &        & $V$ & $2\times240 + 1\times30$ & $1.09-1.07$ \\[6pt]
 OC99  & 08\, 51\, 22.0 & $+$11\, 49\, 00 & 29 Nov & $B$ & $2\times240 + 1\times30$ & $1.36-1.21$ \\ 
 (M~67)   &                &                 &        & $V$ & $2\times240 + 1\times30$ & $1.02-0.98$ \\[3pt]
\hline
\hline
\label{T_log}
\end{tabular}
\end{center}
\end{table*}

\begin{figure*}
\centering
\caption{The $V$ band  of the SMC~5 field covering 
a field of view of $34\times33$ arcmin. This image is the combination
of the two DEEP dithered images. The small black areas seen along the
central part of the image and at the edges are the residuals of the
inter-chip gaps. In this field the following clusters are present:
NGC~346 (the brightest HII region in the SMC), NGC~330, IC~1611,
NGC~306, NGC~299, OGLE~109, OGLE~119, OGLE~99.}
\label{F_smc05}
\end{figure*}

\section{Observations}
\label{S_observation}

The observations for the PF Survey are being carried out using
the WFI camera at the Cassegrain focus of the MPG/ES0 2.2~m telescope
at the La~Silla observatory. WFI is a focal reducer-type mosaic camera
with $4\times2$ CCD chips of $2048 \times 4098$ pixels. Each chip
covers a field of view of $8.12 \times 16.25$ arcmin with a projected
pixel size of $0.238$~arcsec.  The $8$ CCDs are physically separated
by gaps of width $23.8$ and $14.3$~arcsec along the right ascension
and declination directions, respectively.  The full field of view of
the camera is thus $34\times33$~arcmin, with a filling factor of
$95.9\%$. More details about the instrument can be found on the La
Silla web page. 

The adopted observing strategy for the PF survey was a compromise
between number of required passbands, number of fields, exposure-time,
filling factor, and observing overheads.  The observations of PF
fields were conducted in $B$, $V$ and $I$ to provide color information
for the selection of objects.  The integration time was split into a
short-exposure of 30 seconds (SHALLOW), to avoid saturating bright
objects, and two deeper exposures of four minutes each (DEEP). These
were dithered by 30~arcsec both in right ascension and declination.
The long exposures were sufficiently deep to reach the required
signal-to-noise at the spectroscopic limit of FLAMES, while the short
exposures allows one to recover saturated bright stars in the long
exposures exposures (a gain of $\sim4$~mag). Bright stars are also
necessary as guide stars for each FLAMES field, which should be in the
same astrometric system as that of the target list.

Table~\ref{T_log} shows the log of  the observations conducted in  the
period November  $27-29$, $2000$.  The table  gives: in column (1) the
EIS target identification and  the  name of  the primary  object being
observed;  in  columns  (2) and (3)  the   J2000 right   ascension and
declination; in column (4) the date  of observation; in column (5) the
filters used; in column (6) the exposure  time and number of exposures
for the SHALLOW   and DEEP exposures; and  in  column (7)   the seeing
during  the  SHALLOW and DEEP    exposures, respectively. As mentioned
above the $I$ band images are not presented in this paper.

\section{Data Reduction}
\label{S_process}

The WFI images were processed using the EIS pipeline described in more
detail by Arnouts \etal\ (\cite{arnout01}) and Vandame \etal\
(\cite{vandame01}).  The basic reductions steps performed by the
pipeline include the removal of the instrument signature (trimming,
de-biasing and flat-field correction), pixel registration, removal of
cosmic rays and other image artifacts, image swarping to a fixed
astrometric grid and stacking of the swarped images. However, the
pipeline was originally designed to deal with high-galactic latitude
fields and for a completely different observing strategy than the one
adopted for the present survey.  In particular, the fact that the
number of dithered images is small and the fields are crowded required
some changes in the reduction strategy. As discussed below the results
also showed the need for additional modifications of the software,
which are currently underway. One of the issues not addressed in the
present paper is the procedure required for de-fringing the $I$-band
images for which the solution adopted for the Deep Survey is not
applicable to the present observations. It is important to emphasize
that the $I$-filter in use is very red compared to the standard Cousins,
with a broad spectral coverage to the red.  Tests have also shown the
fringing pattern to be variable during the course of one
night. Methods for dealing with the fringing and the impact of its
removal will be discussed in a subsequent paper.

\section{Astrometry}
\label{S_astrometry}

As mentioned above, a key issue of concern regarding the PF survey is
the astrometric calibration of the images and of the target lists
which will be used for FLAMES observations. In this section, the
algorithm used in the calibration of the images and an evaluation of
the available reference catalogs are discussed.

The astrometric calibration performed by the EIS pipeline makes
extensive use of the method developed by Djamdji \etal (1993) based on
the multi-resolution decomposition of images using wavelet
transforms. As described in Arnouts \etal (\cite{arnout01}) this
package is used both to obtain a crude first estimate of a suitable
reference pixel for the WFI images for the run, and an accurate
determination of the astrometric solution for each image. Once an
adequate reference pixel is available for the run the science images
are decomposed into images of different resolutions. The same is done
for a mock image created from the positions of stars taken from the
reference catalog being used.  The lowest resolution images obtained
are then correlated and a first approximation of the astrometric
solution , in this case a translation, is determined.  This solution
is then used to correct the positions of the source catalog extracted
from the next higher resolution science image. The source catalog is
extracted using a simple algorithm which identified local maxima.
This procedure is then repeated for each subsequent resolution while
the search radius used to identify matching objects decreased . At
the highest resolution a polynomial of second-order is used and the
search radius is half a pixel, which in the case of WFI corresponds to
$\sim0.14$~arcsec.

Once an astrometric solution is found for each CCD of the mosaic, the
image is then corrected for distortions by swarping it using a {\em
nearest neighbor} criterion to relocate the flux.  While this proved
to be adequate for the DPS survey, the results of tests discussed
below show that this strategy has to be modified for the PF
survey due to the small number of images per pointing. In this case it
is necessary to re-sample the images by factors of two and three to
obtain the best results.  This re-sampling strategy has been adopted
as a temporary solution. A more general swarping technique is
currently being implemented into the pipeline which will allow for a
suite of kernels to be chosen. This technique and their impact on the
astrometric calibration will be discussed in a subsequent paper of
this series.

After the registration of the individual images a final combined image
can be produced by simply adding the two images using the weight image
to eliminate cosmic rays and bad pixels.  As an illustration,
Figure~\ref{F_smc05} shows the final $V$-band image of the SMC~5 field
covering $34\times33$ arcmin. The observing strategy adopted leads to
a filling factor of $99.91\%$ , albeit with regions less sensitive
than others. Note, however, that the imprints of the the gaps are not
completely removed and can still be seen aligned along the central
part of the image and at the edges.  It is interesting to note the
large number of stellar systems that can be seen in this field, among
them: NGC~346, NGC~330, IC~1611, NGC~306, NGC~299, OGLE~109, OGLE~119,
OGLE~99.

Considering the importance of an accurate astrometric solution in the
preparation of target lists for any fiber system, such as FLAMES,
several tests were performed in order to evaluate and fine-tune the
"swarp" algorithm and to chose the best reference catalog.  These
tests were conducted using the data for the nearby open cluster M~67
(EIS target OC99), for which results of several previous work are
available in the literature. For instance, the work of Girard \etal\
(\cite{gira89}) provides a source catalog with positions having a
typical internal relative error of $\simeq0.021$~arcsec for a sample
of $663$ stars, with errors of $\simeq0.010$~arcsec for the stars with
$V<14$.  Their estimated external (absolute) error is $0.16$ arcsec.
The catalog also gives proper motions measurements for each star for a
mean reference epoch of $1950.8$.

\void {Also of interest is the work of Montgomery \etal
(\cite{mont93}) which provides a photometric catalog that is used
below to check the photometric catalog produced in this paper.}

\begin{figure}
\centering
\caption{Comparison of the  two independent astrometric calibration of
the SHALLOW and DEEP  exposures adopting  the  USNO catalogue for  the
field of  M 67 (OC99).    {\em (A)}  Nearest neighbor algorithm   with
sub-sampling of $1\times1$.   {\em (B)} same as  above but  with pixel
sub-sampling of $2\times2$.   {\em (C)} same as  above but with  pixel
sub-sampling of $3\times3$.The   vertical and horizontal  dashed lines
mark the mean  residuals in  RA and Dec   after applying  a  $3\sigma$
clipping to the data.  The mean values  with the final $1\times\sigma$
rms are given on the  figure.  Also shown are   the histograms of  the
residuals as a function of RA (top) and Dec (right).}
\label{fig_subsam}
\end{figure}

\begin{figure}
\centering
\caption{The two-dimensional (RA, Dec) distribution of positional 
residuals   obtained  from the   comparison   of the   astrometrically
calibrated  SHALLOW  catalogs  of  M~67,  using  different   reference
catalogs, and the  reference  catalog  used: USNO~2.0  (upper  panel);
GSC~2.1 (middle  panel); and GSC~2.2  (lower panel).  The vertical and
horizontal dashed  lines mark the mean residuals  in RA and  Dec after
applying a  $3\sigma$ clipping to the data.   The mean values with the
final $1\times\sigma$ rms are given in the figure.  Also shown are the
histograms of the residuals as a function of RA (top) and Dec (right).}
\label{fig_radec}
\end{figure}

\subsection{Fine-tuning the pipeline}
\label{S_fine-tune}

A first set of tests were carried out to establish the best
re-sampling strategy to be used in the astrometric calibration of the
images.  First, the SHALLOW and DEEP images of OC99 were
astrometrically calibrated using the nearest neighbor algorithm
without image re-sampling, the default mode of the pipeline. Second,
the photometric pipeline (see Section~\ref{S_photo}) was run to
extract point-like objects and construct catalogs for both the SHALLOW
and the combined DEEP images.  The reader is reminded that the SHALLOW
and DEEP images were astrometrically calibrated separately and thus
have independent solutions.  The upper panel of
Figure~\ref{fig_subsam} (panel A) shows a comparison between the
derived SHALLOW and DEEP catalogs.  From the figure one can clearly
see the undesirable effect of the nearest neighbor approach. In this
case, the distribution of the positional residuals in RA and Dec are
highly non-Gaussian, with the residuals in RA being described by a
bi-modal distribution with the peaks separated by $\sim0.24$ arcsec,
corresponding to one WFI pixel.  This systematic effect is an artifact
introduced by the nearest neighbor approach without image
re-sampling. This effect becomes obvious when one is dealing with a
small number of images, as in the present case.  Better results are
obtained when the image is re-sampled and each original pixel is split
into $2\times2$ pixels.  This case is shown in the middle panel of
Figure~\ref{fig_subsam} (panel B), where the peaks of the bi-modal
distribution are much closer together and the distribution of the
residuals resembles more closely that of a Gaussian. Even better
results are obtained using a higher resolution re-sampling with each
pixel being split into $3\times3$ pixels. This is shown in the lower
panel of the same figure (panel C). The latter sampling leads to a
Gaussian distribution with an $rms$ of $\simeq0.05$~arcsec in both
coordinates.  For all three cases the variation of the RA and Dec
residuals as a function of magnitude were inspected and were found to
be negligible.  Note that while the mean offset of the residuals do
not change the $rms$ is significantly smaller in the right ascension
direction between the first and last cases.

The results of these tests show that in order to avoid systematic
errors between bright and faint sources a re-sampling of the image is
required if the nearest neighbor strategy is adopted.  In this paper
all images have been re-sampled by splitting the original pixels in
$3\times3$ sub-pixels. However, since re-sampling the image is costly
in processing time, a more general swarping algorithm has been
developed to avoid the discreteness effects of the nearest neighbor
approach. This algorithm is currently being implemented into the EIS
pipeline. It will be used for all subsequent reductions of the PF
survey data.

\begin{figure}
\centering
\caption{
The positional residuals in RA and Dec computed as in
Figure~\ref{fig_radec}, are shown as a function of RA and Dec of the
EIS catalog positions for the USNO~2.0 (upper panel); GSC~2.1 (middle
panel); and GSC~2.2 (lower panel).}
\label{fig_cata_radec}
\end{figure}

\begin{figure}
\centering
\caption{
The positional residuals in RA and Dec computed as in
Figure~\ref{fig_radec}, shown as a function of the magnitude listed in
the respective reference catalogs used for the USNO~2.0 (upper panel);
GSC~2.1 (middle panel); and GSC~2.2 (lower panel). Points represented
by filled circles were used in the linear fit to the data.  The fit
parameters are listed in Table~\ref{fig_cata_rms}.}
\label{fig_cata_mag}
\end{figure}

\subsection{Choice of the reference catalog}
\label{S_what_catalog}

Another important issue to consider is the impact of the reference
catalog used. In order to evaluate the available catalogs, the
amplitude of the residuals, their dependence on position and on
apparent magnitude, and their comparison with an independent
astrometric catalog are investigated in this section.  It is important
to point out that during the analysis of the data presented in this
paper only two reference catalogs were available - the USNO~2.0 and
the GSC~2.1, a pre-release version of the Guide Star Catalog
version~2, used as the reference catalog in the most recent EIS
reductions (\eg Vandame \etal\ 2001; Arnouts \etal\ 2001). However,
right before the release of these data, the GSC~2.2 version (McLean
\etal\ 2001) became available. While it was not possible to re-analyze
all  of the data in  time for the  scheduled release,  the GSC~2.2 was
used in the tests  presented here to assess  the differences, if  any,
relative to the pre-release version. This  is important because at the
time of writing a number of fields observed as  part of the PF
were not yet covered by the GSC~2.2. This  raised the issue of whether
the GSC~2.1 or  the USNO~2.0 should be  used for the  first PF
release.

In the present analysis the SHALLOW images of M~67 were calibrated
using the three reference catalogs described above.  Next, the
positional differences between the calibrated source list and the
reference catalog were computed. The RA and Dec residuals are shown in
Figure~\ref{fig_radec} for each reference used.  For consistency, only
objects brighter than $m=18.5$ are plotted, this lower limit which
corresponds to the magnitude cutoff of the GSC~2.2 catalog.  From the
figure one can immediately see the superiority of the GSC~2.2 relative
to the other catalogs considered.  The distribution of the residuals
using the GSC~2.2 catalog is highly concentrated and Gaussian in both
directions. This is in marked contrast to the other two which have a
much wider spread in the $\Delta$RA-$\Delta$Dec plane, and either very
skewed (USNO~2.0) or bi-modal (GSC~2.1) distributions.  The mean
offsets and the $rms$ of the distributions, given on each panel, show
that while the systematics of the USNO~2.0 and GSC~2.1 may be
different, which could be partly due to the different epoch of the
photographic plates, both have comparable accuracy
($\sim0.25$~arcsec), as measured by the $rms$ of the distribution.  On
the other hand, GSC~2.2 represents a real improvement with the $rms$
of the residuals being $\lsim0.12$~arcsec.

To understand the nature of the observed spread of the residuals, the
$\Delta$~RA and $\Delta$~Dec residuals are plotted as a function of RA
and Dec taken from the calibrated EIS catalogs.  This is shown in
Figure~\ref{fig_cata_radec} and for each of the considered catalogs.
In general, besides the large scatter, no noticeable systematics are
seen. Note, however, the two bands of data points seen in the GSC~2.1
residuals, the reason for which is discussed below. It is worth
pointing out that while no correlation between the residuals and
position were found, other than that previously mentioned, for the
particular field considered here, GSC~2.1 did show a systematic
variation of the residuals as a function of both RA and Dec in the
case of the Chandra Deep Field south located at the edge of the plate
( Arnouts\etal\ 2001). It remains to be seen if this effect has been
corrected for in the GSC~2.2.

Complementing the above plots Figure~\ref{fig_cata_mag} shows the
variation of the right ascension and declination residuals as a
function of magnitude for the USNO~2.0 (top panel), GSC~2.1 (middle
panel) and GSC~2.2 (bottom panel). Also shown is the linear fit to the
residuals as function of magnitude. The fit parameters are given in
Table~\ref{T_catalogs_fits}.  From the figures and the table one finds
a relative strong dependence of the residuals on the magnitude for the
USNO~2.0 and a striking discontinuity at $m\sim15$ in the
GSC~2.1. Only GSC~2.2 shows no systematic behavior with
magnitude. From the liner fit of the USNO residuals one finds an
offset in the RA residuals of $\sim0.30$~arcsec between bright and
faint stars ($12<V<18.5$). The effect in Dec computed over the same
magnitude interval is negligible. This systematic error is large
compared to the required accuracy of the relative positions. This
offset is significantly smaller in the case of GSC~2.2 being
$\lsim~0.10$, in both directions. The systematic variation of the
residuals with the magnitude explains the amorphous distribution seen
in Figure~\ref{fig_radec}. This magnitude dependence has no bearing on
the astrometric calibration of the pipeline. Note that this means that
one should not use the coordinates of a reference star as listed in
the reference catalog. Instead, stars used for pre-setting the
telescope must share the same astrometric system as the objects being
observed.

\begin{table}[h]
\caption{Linear fit parameters for the positional residuals
as a function  magnitude.}
\begin{tabular}{llrrl}
\hline\hline
Catalog & Residual & {\tt slope} & {\tt zero-point} & {\tt rms} \\[3pt]
\hline
&&\\[-9pt]
USNO   & $\Delta$  RA  & -0.046 & 0.684  & 0.262 \\
USNO   & $\Delta$  Dec & 0.011  & -0.149 & 0.232 \\
GSC~2.1\tnote{*}& $\Delta$  RA  & -0.009 & 0.388  & 0.180\\
GSC~2.1& $\Delta$  Dec & -0.019 & 0.248  & 0.205 \\
GSC~2.2& $\Delta$  RA  &  0.014 & -0.268 & 0.128 \\
GSC~2.2& $\Delta$  Dec & -0.024  & 0.381 & 0.139 \\[3pt]
\hline\hline
\label{T_catalogs_fits}
\end{tabular}
\begin{tablenotes}
\item[*] Only objects brighter than $14.8$
\end{tablenotes}
\end{table}

\begin{figure}
\centering
\caption{
The two-dimensional (RA, Dec) distribution of positional residuals
obtained from the comparison of the astrometrically calibrated
catalogs of M~67, using different reference catalogs, with the
astrometric catalog of Girard \etal\ (\cite{gira89}). The EIS catalogs
were calibrated astrometrically adopting: USNO~2.0 (upper panel);
GSC~2.1 (middle panel); and GSC~2.2 (lower panel).  The vertical and
horizontal dashed lines mark the mean residuals in RA and Dec after
applying a $3\sigma$ clipping to the data.  The mean values with the
final $1\times\sigma$ rms are given in the figure.  Also shown are the
histograms of the residuals as a function of RA (top) and Dec (right).
}
\label{fig_cata_rms}
\end{figure}

As an external check of the different astrometric solutions obtained
for M~67 these were compared with the astrometric catalog of Girard
\etal\ (\cite{gira89}).  In  this comparison the position of the stars
 in the Girard \etal\ catalog were corrected for proper motion to the
PF epoch of observation.  The distribution of the positional
residuals in the RA-Dec plane are shown in Figure~\ref{fig_cata_rms}
for the EIS catalog produced for M~67 field using USNO~2.0 (top panel),
GSC~2.1 (middle panel) and GSC~2.2 (bottom panel). Also shown in the
figure are the distributions of the residuals in the RA and Dec
directions. In each panel the mean offset and $rms$, computed after
applying a $3\sigma$ clipping to the data, are also shown. While the
$rms$ values obtained are comparable ($\lsim0.15$~arcsec), only for
GSC~2.2 are the residuals well represented by a Gaussian distribution.
Note that the mean offsets are not relevant for the preparation of
target lists for FLAMES since its fiber positioner is allowed to move
within a $2$~arcsec window (Pasquini, private communication).

In conclusion, an analysis of the three reference catalogs currently
available for the astrometric calibration of the WFI images shows that
the recently released GSC~2.2 catalog yields the best results overall,
with less systematics than the USNO~2.0 current in use by EIS. The
analysis also reveals some unexpected features of the GSC~2.1 used in
the calibration of the Chandra Deep Field south (EIS DEEP~2c field;
Arnouts \etal 2001), pointing out the need for a re-evaluation of the
astrometric calibration of those images. Nevertheless, it is important
to emphasize that in all cases the $rms$ of the residuals are below
the $0.2$~arcsec upper-limit required for the optimal performance of
FLAMES. Since at the time of writing the GSC~2.2 catalog did not cover
the SMC fields presented in this paper, the USNO~2.0 was used as the
reference catalog. However, the results presented in this section
clearly suggest that the GSC~2.2 reference catalog shows no
significant systematic errors and is currently the best
available. Future EIS calibrations will be done using this catalog, as
soon as all the fields of interest are available. The results also
demonstrate that the algorithm being employed in the astrometric
calibration of the WFI frames in the EIS pipeline is robust, yielding
results within the requirements of FLAMES. One also finds that the
accuracy of the astrometric solution is limited by the intrinsic
errors of the reference catalog used.

\section{Photometry}
\label{S_photo}

\subsection{Source Extraction}

Source extraction and stellar  photometry (PSF fitting technique) were
performed using the  DAOPHOT/ALLSTAR package (Stetson  \cite{stet87}).
Minor changes to these programs had to be made to  enable them to read
fits images directly   and  to  handle $8$k$\times8$k  images,   which
basically  required  increasing the  memory buffer,   the size  of the
buffer used for determining the sky level and the number of stars used
to model the  PSF  function for  a  frame. In  addition, a  number  of
procedures were implemented to automatize the whole procedure and make
it more adaptable to the pipeline framework.

The first step is to create a model PSF for each image.  The building
of the PSF for each frame was done iteratively using a set of isolated
and unsaturated stars.  A preliminary list typically containing $500$
stars is extracted and a first rough PSF is generated.  Objects with
poor fits are discarded from the list and a new PSF is created.  This
process is repeated until the fitting solution is acceptable for all
objects. Typically, the process converges after a few iterations with
the final PSF determined by at least $300$ stars per image.  The final
model for the PSF is generated with a ``Penny'' function which has a
quadratic radial dependence on the stellar coordinates.  The resulting
PSF is then used by the ALLSTAR routine to measure the magnitudes of
all point sources detected by the FIND command of DAOPHOT.  The final
star-subtracted image is then visually inspected to check the result.

The above procedure was used to produce stellar catalogs for the
SHALLOW and DEEP images of all selected fields in $B$ and $V$.  The
catalogs extracted from the combined DEEP images are very sensitive to
the detection threshold used.  In all cases the detection threshold
was set to $3\sigma$ above the background counts, the best compromise
between completeness and the number of spurious detections. Lower
thresholds cause a large increase in the number of spurious
objects. This is especially true for the DEEP images which have a
lower signal-to-noise at the edges and over regions sampled by a
single image due to the inter-chip gaps. This means that some faint
objects within these regions might be missed. Information regarding
the location of these regions can be found in the corresponding
weight-maps (see section~\ref{S_release}). 

Since the DAOPHOT/ALLSTAR package only works in pixel coordinates, the
astrometric calibration available in the header of the image being
processed is used to transform the resulting catalog to the
world-coordinate system ($wcs$).  This is in contrast to the procedure
adopted by the EIS pipeline, which normally uses SExtractor which
supports the $wcs$ convention. Once catalogs from the SHALLOW and DEEP
images are available those in the same passband are matched and the
magnitude of objects in the SHALLOW catalogs are scaled to those in
the DEEP catalogs. In the process the error weighted mean of the
magnitude differences magnitudes is computed for the stars in common,
which never exceed $0.03$ mag. It is important to point out that the
SHALLOW and DEEP images overlap over an interval of $6$~mag. The
SHALLOW and DEEP catalogs are then merged and only objects satisfying
the conditions $CHI<1.2$ and $-1.2<SHARP<1.2$ are kept in the final
combined catalog. The parameters $CHI$ and $SHARP$ are a measure of
the quality of the PSF fitting and the sharpness of the light profile
of the object, respectively. Their exact definition can be found in
the documentation of the DAOPHOT/ALLSTAR package.  Finally, a $BV$
color catalog is produced by associating the entries of the combined
single passband catalogs described above, using a matching radius of
0.8~arcsec. The color catalogs also include objects detected in only
one passband to avoid discarding intrinsically ``blue'' or ``red''
stars.  The instrumental PSF magnitudes are then aperture-corrected
and calibrated as described in the next section.

\subsection{Photometric Calibration}

The images were calibrated to the Johnson-Cousins system using
observations of Landolt (\cite{land92}) standard stars. Photometric
solutions were determined using mean extinction coefficients and
empirically determined color terms given by the equations:

\begin{equation}
B_{JC}=B_{EIS-842} + 0.29 \times (B-V)_{JC} 
\end{equation}
\begin{equation}
V_{JC}=V_{EIS-843} - 0.08 \times (B-V)_{JC} 
\end{equation}

\noindent where $B_{EIS-842}$ and $V_{EIS-843}$ are in the EIS
magnitude system, and the numbers correspond to ESO's filter
identification number.  The extinction coefficients for each passband
were computed from the mean extinction curve for La Silla and the
response of the WFI filters used yielding: $K_{B}=0.235$ and
$K_{V}=0.135$.

The instrumental PSF magnitudes were converted into aperture
magnitudes, assuming that $m_{ap}=m_{PSF}+constant$ (Stetson
\cite{stet87}), where the $constant$ is the aperture correction. This
was obtained for each image from the analysis of the growth curves of
bright isolated stars identified on each frame. Typical values for the
aperture correction are in the range $0.15-0.3$ magnitudes.  In the
case of crowded regions like the SMC fields aperture corrections were
calculated after subtracting all stars around the measured bright
objects.

\begin{figure} 
\centering
\caption{The M~67 (OC99) color-magnitude diagram obtained from the
final combination of the SHALLOW and DEEP data.}
\label{F_cmdbv_m67}
\end{figure}

As an example of the application of the above procedure
Figure~\ref{F_cmdbv_m67} shows the color-magnitude diagram obtained
for the SHALLOW (top panel) and the DEEP (bottom panel) color
catalogs.  Note that the color-magnitude diagram derived from the
SHALLOW catalog extends from $V~10$ to $V\sim20$, while that derived
from the DEEP catalog covers the magnitude interval $14
\lsim V \lsim 23$.  The color-magnitude diagram of  the final merged
catalog covers an impressive 13~mag interval, as shown in
Figures~\ref{F_cmd_all1} and~\ref{F_cmd_all2}  .

The number of objects given in the final $BV$ color catalog for each
field is listed in Table~\ref{T_number}. The table gives: in column
(1) the name of the primary target; in column (2) the EIS
identification name; and (3) the total number of sources.

\begin{table}[h]
\caption{Total number of sources in the released $BV$ catalogs.}
\begin{tabular}{llr}
\hline\hline
Name & EIS name & Number \\[3pt]
\hline
&&\\[-9pt]
Berkeley~20 & OC03 &   7190 \\
NGC~2477    & OC12 &  38800 \\
NGC~2506    & OC14 &  18900 \\
Messier~67  & OC99 &   4290 \\
SMC         & SMC~5& 280000 \\
SMC       & SMC~6& 246000 \\[3pt]
\hline\hline
\label{T_number}
\end{tabular}
\end{table}

\begin{figure}
\centering
\caption{Magnitude and color comparison of the EIS catalog with other
authors. The {\em upper panel} shows a comparison of the EIS
photometry of the OC~99 field with that of Montgomery \etal\
(\cite{mont93}).  The the $(B-V)$ and $V$ residuals are plotted
against EIS $V$ calibrated magnitudes.  The {\em lower panel} shows a
similar comparison but for NGC~330 (SMC~5) with the Vallenari \etal\
(\cite{valle94}). In these plots the horizontal lines indicate the
median value of the differences.  }
\label{F_m67_montgomery}
\end{figure}

\subsection {Comparison with Other Authors}

As  a  check of the  photometric  catalogs being  produced,  these are
compared to the  data from Montgomery  \etal\ (\cite{mont93}) for M~67
(OC99) and Vallenari \etal\ (1994)  for NGC~330 (SMC5). The results of
these comparisons are  shown in Figure~\ref{F_m67_montgomery}. The two
sets of panels refer to M~67 and NGC~330, respectively. In these plots
the  differences in the   $(B-V)$ color (upper   plot) and in  the $V$
magnitude  (lower  plot) are displayed as   a function of the  EIS $V$
Johnson-Cousins magnitude. In these    plots all stars in   common are
included. Circles indicate those used in computing the mean offset and
$rms$ and crosses those discarded by  applying a $3\sigma$ clipping to
the data.

For M~67 one finds a mean offset in the photometric zero-point of
$0.068$~mag and an $rms$ of $ 0.064$~mag in the magnitude interval
$10<V<20$. For the $(B-V)$ color one finds a mean offset of
$-0.017$~mag and an $rms$ of $0.11$~mag. In the case of NGC~330 one
finds a mean value of magnitude differences of $0.050$~mag, similar to
the one mentioned above, and an $rms$ of $0.18$~mag, while the mean
offset in $(B-V)$ is $0.053$~mag with an $rms$ of $0.15$~mag, in the
same magnitude interval as for M~67. Together these results show that
the measured colors are in excellent agreement with those measured by
other authors despite of the large color term involved in the
correction of the WFI instrumental magnitudes to the Johnson-Cousins
system. The data also show that the present data may have a systematic
error of $\lsim0.07$~mag in the zero-point which should be further
investigated. More importantly, if one assumes that the random errors
for EIS and the Montgomery \etal data are comparable then the measured
$rms$ of the magnitude differences for M~67 implies an error of
$\sim~0.04$~mag at $V\sim20$ in the EIS catalogs. This value is
consistent with the internal estimates and the original requirements
for the PF survey.  The larger scatter measured from the comparison of
NGC~330 may be due to the larger uncertainties in the magnitudes due
de-blending problems in such very dense system.

\section{Survey Products}
\label{S_release}

The images already released include the combined deep $B$ and $V$
exposures of each field, with all images normalized to $1$ second
exposure, and are presented in the TAN projection. In the data release
the science images have been combined with their corresponding
weight-maps into a single fits file containing two image extensions.

In addition to the pixel maps, for each of the six observed fields
discussed in this paper, the following catalogs are also available:

\begin{enumerate}

\item Three instrumental photometric catalogs for each passband: the
SHALLOW, the DEEP and the combined catalog, all in ASCII format. The
following naming convention is adopted

\medskip
FIELD\_FX\_N\_YYYY-MM-DDTHH:MM:SS.asc
\medskip

\noindent where FIELD is EIS target name, FX
is the passband, N varies from 1 to 3  and corresponds to the SHALLOW,
DEEP and COMBINED   catalogs, respectively.  The remaining  characters
refer to the date and time of the catalog production.

\item A calibrated $BV$ color catalog  available in three
different formats for the convenience of the interested user: a FLAMES
input file {\tt .fld}, a SKYCAT input file {\tt .scat}, and a normal
ASCII file {\tt .asc}. The name of each file follows the convention
described above.

\end{enumerate}

\begin{table*}[t]
\begin{center}
\caption{An example of a FLAMES operating system input file (.fld)}
\begin{tabular}{r|lccccccl} 
\hline\hline
&\multicolumn{8}{l}{LABEL EIS PF FIELD OC99} \\
&\multicolumn{8}{l}{UTDATE 2001 07 01}\\[-6pt] Header &&&&&&&&
\\[-6pt] &\multicolumn{7}{l}{CENTER 08 51 23.073 11 50 7.83}\\
&\multicolumn{7}{l}{*EQUINOX J2000.0}\\
\hline
Targets &&&&&&&& \\[-6pt]
&OC99\_000001 &  08 50 26.850 &  11 33 43.03 & P & 1 & 16.620 & 0 &  1.663\\
&OC99\_000002 &  08 52 04.271 &  11 33 49.17 & P & 1 & 18.292 & 0 &  1.509\\
&OC99\_000003 &  08 51 01.327 &  11 33 50.11 & P & 1 & 19.078 & 0 &  0.829\\
&OC99\_000004 &  08 51 10.555 &  11 33 53.43 & P & 1 & 17.743 & 0 &  1.219\\
&OC99\_000005 &  08 51 40.304 &  11 33 55.29 & P & 1 & 17.299 & 0 &  0.885\\
&OC99\_000006 &  08 50 19.500 &  11 33 58.38 & P & 1 & 19.069 & 0 &  1.460\\
&OC99\_000007 &  08 50 26.126 &  11 33 59.22 & P & 1 & 18.705 & 0 &  1.049\\
&OC99\_000008 &  08 50 28.017 &  11 34 01.92 & P & 1 & 20.661 & 0 &  1.506\\
&OC99\_000009 &  08 52 05.010 &  11 34 08.15 & P & 1 & 18.423 & 0 &  1.515\\
\hline\hline
\label{T_FLAMES}
\end{tabular}
\end{center}
\end{table*}

An example of the FLAMES input {\tt .fld} file is shown in
Table~\ref{T_FLAMES}.  The first four rows of the file are part of a
header which includes: a LABEL with the name of the field; the UTDATE
of the observations (this have to be changed according to the needs of
the user); the CENTER of the field in RA and Dec coordinates; the
EQUINOX of the coordinates (in this case, J2000.0). The targets
following the header have the following fields: in column ($1$) the
identification sequential number; in columns ($2$)$-$ ($4$) the RA of
the object in hours; in columns ($5$)$-$($7$) the declination of the
object in degrees; in column ($8$) one letter specifying the kind of
fibers to be used among the different types available (e.g. FLAMES in
MEDUSA mode is be identified with the "P" letter); in column ($9$)
observation priority of the objects (highest $1$; lowest $9$), in this
example $1$ is given for all the object; in column ($10$) measured $V$
magnitude of the object; in column ($11$) another identification code
($0$ in this example); and in column ($12$) a comment field, which in
this example lists the $(B-V)$ color.

\section{Discussion}
\label{S_discusion}

\subsection {Color-Magnitude Diagrams}

In order to illustrate the results obtained  from the photometric data
presented here,  Figures~\ref{F_cmd_all1}  and  ~\ref{F_cmd_all2} show
the  color-magnitude diagram  (CMD)  for each  of the observed fields.
For most  fields the CMDs were obtained  using the  catalog within the
entire   area covered by   WFI.  However,  and   in order  to minimize
foreground/background contamination, for the open clusters Berkeley~20
and  NGC~2506, the CMDs shown  correspond  to a circular region around
the nominal cluster center of 3 and 5~arcmin in radius, respectively.

Even though still a small sample, the examples presented here show the
large variety of stellar systems being observed by the PF
survey in terms of age, metallicity, size, distance and
environment. The wide-area and the extended magnitude coverage
($\sim13$~mag) down to $V\sim23$ provide an invaluable data set to
extract samples suitable for the scientific drivers of FLAMES which
include, among others, studies of: chemical abundances of stars in
clusters and selected galactic regions (Bulge, Disk, and Halo);
stellar kinematics and structure of stellar clusters; chemical
composition and dynamics of nearby dwarf spheroidal galaxies;
circumstellar activity in young stellar objects; very low mass stars
and brown dwarfs in star forming regions. 

From  Figures~\ref{F_cmd_all1}  and   ~\ref{F_cmd_all2} one   can  see
systems  with   well-defined  main-sequences (MS),    probable  binary
sequences, blue   straggler populations,   red clump  stars, potential
white dwarf candidates,  very red objects  and systems with  composite
stellar populations  including very   young stellar associations,  and
additional information about the galactic structure.  These points are
briefly discussed on an individual basis below.


\subsection {Comments on Individual Systems}

\begin{figure*}
\centering
\caption{The CMD for M~67, Berkeley~20 and NGC~2477, 
obtained from the   combination of  the  catalogs  extracted  from the
SHALLOW and  DEEP images, as     described in the text.  To   minimize
foreground contamination   the   CMDs of  Berkeley~20    corresponds to
circular regions of $3$~arcmin in  radius around the nominal center of
the cluster.}
\label{F_cmd_all1}
\end{figure*}

\begin{figure*}
\centering	
\caption{The CMD for NGC~2506, SMC~5 and SMC~6, 
obtained from the   combination of  the  catalogs  extracted  from the
SHALLOW and  DEEP images, as     described in the text.  To   minimize
foreground contamination   the   CMDs of  NGC~2506    corresponds to
circular regions of $5$~arcmin in  radius around the nominal center of
the cluster.}
\label{F_cmd_all2}
\end{figure*}

\subsubsection{M~67}

The close location to the Sun ($(m-M_V)=9.59$), the low reddening,
($E_{B-V}=0.05$; Montgomery \etal\ \cite{mont93}), the rich population
($800$ M$_\odot$; Mathieu \cite{math85}; Montgomery
\etal\ \cite{mont93})  and the minimum contamination by foreground/background
stars make of M~67 one of the best studied open clusters. By fitting
theoretical isochrones to the observed MS Dinescu \etal\
(\cite{dine95}) derived an age of $4.0\pm0.5$ Gyr.  Tautvaisiene
\etal\ (\cite{taut00}) estimated a mean metallicity of
[Fe/H]$=-0.03\pm0.03$. This system is also of great interest because
of its rich population of blue-stragglers, X-ray sources and white
dwarfs. In many ways M~67 is an ideal case for studies of the
structure and evolution of Population~I stars and thus a prime target
for a spectroscopic follow-up with FLAMES. 
Moreover,  and thanks to  the  precise proper-motions  study  of $663$
stars by Girard \etal\  (\cite{gira89}), M~67 has been  used as a test
case  for  the astrometric   calibration   procedures adopted in   the
reduction of the PF data (Section~\ref{S_astrometry}).

From the  CMD shown   in   Figure~\ref{F_cmd_all1} the  MS  is  clearly
discernible  from the cluster  turnoff  (TO) at $V\simeq12.8$ down  to
$V\simeq22$.  At $V\sim18.5$  the MS shows  an abrupt change in slope.
At fainter magnitudes the $(B-V)$ is almost constant ($\simeq1.8$) due
to the saturation of this color for  very cool stars, corresponding to
M$\sim0.5$~M$_\odot$.    This  may  be   also    caused by a    strong
contamination by background stars and  galaxies at the very faint  end
of the MS.  The large magnitude baseline  of the CMD allows allows one
to probe the  red giant  branch  up to $V\sim10.0$.  One  can also see
several blue  stragglers at magnitudes brighter  than $V=12.5$  and in
the color interval $-0.6<(B-V)<0.8$.

Close inspection of the CMD suggests the  presence of a sparse, redder
sequence running  parallel to the  cluster main  sequence, as  seen by
Montgomery  \etal  using $BVI$   data (see also Section~\ref{future}).
This has  been  interpreted as  due  to  binary systems,  estimated to
constitute $38\%$ of the cluster  population. Furthermore, Pasquini \&
Belloni (\cite{pasq98})  based   on photometric  and   high-resolution
spectroscopic studies of  ROSAT   sources concluded that most  of  the
detected X-ray sources are binaries. These ROSAT sources have peculiar
locations in the cluster CMD consistent with the above interpretation.

Finally, several blue objects ($(B-V)<0.4$) are visible fainter than
$V\simeq19$. These objects could be the brightest members of the white
dwarf (WD) cooling sequence found by Richer \etal\ (\cite{rich98}), in
the absolute magnitude interval of $M_V=10-14.6$.

\void{They estimated that WDs currently contribute about $9\%$ of the total
cluster mass, and that such number appears to be somewhat low when
compared with the number of giants observed in the cluster.}

\subsubsection{Berkeley~20}

From  the study of  several stellar systems carried  out  by Phelps et
al. (\cite{phelps94}) Berkeley~20 is ranked as one  of the oldest open
cluster, with an age comparable to that of  M~67.  From an analysis of
the  $V-(V-I)$ CMD MacMinn  et al.  (\cite{macminn94}) estimate a mean
age of $6$ Gyr and a metallicity of [Fe/H]$\simeq-0.23$.  However, the
system      is further away,     with      a  distance  modulus     of
(m$-$M)$_V~\simeq15.0$~mag.   It  is also  located  in  a direction of
higher extinction, estimated to be $E_{V-I}\simeq0.16$ mag.  Moreover,
it is very close  to HD~64503,  a  bright star with  $V\sim4.5$.  This
explains  some of the stray light  seen  over the images available for
this  field.   Nevertheless,   both the galactocentric  distance   and
distance  from the plane (MacMinn   et  al. \cite{macminn94}) make  of
Berkeley~20    the most  distant known  open    cluster, and a  better
determination   of  its metallicity  and  age   would  help  trace the
formation and evolution of the galactic disk.

Due to the cluster distance, the field covered by  WFI leads to a high
foreground contamination. In  order  to minimize this effect  the  CMD
shown in   Figure~\ref{F_cmd_all1}   includes only objects    within  a
circular region with  $3$ arcmin radius  around the nominal center  of
this open cluster.  Within this region, the  CMD shows a  well-defined
turnoff at $V\simeq18.25$ and $B-V\simeq0.6$.  There are also a number
of blue-straggler candidates. However, no red clump  is visible in the
CMD shown in  Figure~\ref{F_cmd_all1}.   On   the other hand,  if   one
inspects  the CMD obtained  using a larger  radius ($\sim10$ arcmin) a
small and isolated number of stars are found around $V\simeq16.30$ and
$(B-V)\simeq1.1$.   While at such  distance from the cluster center it
is difficult to  discriminate cluster members  from field stars, it is
interesting  to  note that  the location  of this  concentration would
coincide with the red clump  of M~67 (applying the appropriate scaling
for the  cluster   distance).  This  result  is  consistent with   the
conclusions of MacMinn et al. that the general features of Berkeley~20
CMD   are  similar to those of   M~67,  even though  no  red clump was
identified in their data. This shows the great  value of the wide-area
coverage  of the PF  data.  Follow-up observations  with FLAMES  could
settle this question.

\subsubsection{NGC~2477}

In contrast to the previous systems, NGC~2477 is a relatively young
cluster $1^{+0.3}_{-0.2}$ Gyr located at a distance of $(m-M)_0=
10.61$ (Kassis \etal\ \cite{kassis97}). Unfortunately, as shown by
Kassis \etal\ and Majewski \etal\ (\cite{majewski00}), NGC~2477
suffers from differential reddening, and both groups adopted a
reddening range $E_{B-V}=0.2-0.4$ determined by Hartwick \etal\
(\cite{hartwick72}).  Friel \& Janes (\cite{friel93}) determined a
metal abundance of [Fe/H]$=-0.5\pm0.11$ using moderate resolution
spectroscopy of seven cluster giants. 

The CMD of NGC~2477 shows that the MS of the cluster is well-defined
from $V\simeq12.7$ to $V\simeq22$.  The TO of the cluster is located
around $(B-V)\simeq0.50$, at about the same magnitude of the red clump
($V\simeq12.75$) seen at $(B-V)\sim1.2$.  The effect of the
differential reddening is evident at the TO magnitude, which widens
the MS near the TO. Finally, some blue-stragglers candidates are found
at $(B-V)\simeq0.2$ and $V\lsim12$.

A population of blue objects ($(B-V)\lsim0.6$ is  present
at  $V\gsim18$. However, the density of points in this region is far
less than that seen in the CMD presented by Kassis
\etal\ (\cite{kassis97}) over a smaller area ($15\times15$~square
arcmin). A smaller density is also found by Majewski \etal\
(\cite{majewski00}), even though in a different magnitude
system. Furthermore, the blue population of Kassis \etal\ are spread
over a large range of color and do not seem to define any clear
sequence which could reflect the WD cooling sequence. While the blue
population observed with the present data may be associated with white
dwarfs, as hinted by the presence of few faint stars around $V\sim20$
and $(B-V)\simeq0.2$, the Kassis \etal\ seem to considerably
overestimate the density of these objects. The reasons for that are at
the present time not clear.

One prominent feature of the NGC~2477 CMD is the substantial
contribution of background field stars, which is independent of the
region around the cluster considered. Fortunately, the contamination
is concentrated in a well-defined color range ($0.8<(B-V)<1.6$) at
magnitudes $V\gsim18.0$, not severely affecting the MS of the cluster,
except where it intersects the cluster MS around $V\sim18$.  The
effect of the contamination, by field giant stars,  is also seen at brighter
magnitudes and redder colors, almost overlapping the cluster's red
clump.

\subsubsection{NGC~2506}

This cluster has been studied by McClure \etal\ (\cite{mcclure81})
over a $10$ arcmin diameter field of view, using photoelectric
photometry, and by Marconi \etal\ (\cite{marconi97}), who obtained CCD
photometry of an area of $6\times6$ arcmin.  The present data reach
almost four magnitudes deeper than any previous study of this system,
covering at the same time a much larger area, at least 10 times the
size. NGC~2506 is a relatively old open cluster but its exact age is
very uncertain, varying from $1.5-3.4$ Gyr. Marconi \etal\ estimate
$1.5-1.7$ Gyr, in reasonable agreement with the value of $1.9$ Gyr
found by Carraro \& Chiosi (\cite{carraro94}), while McClure \etal\
(\cite{mcclure81}) estimate an age of $3.4$ Gyr.

For this cluster Figure \ref{F_cmd_all2} shows only objects within a
circular region 5~arcmin in radius.  The CMD of this region shows a
well-defined MS closely resembling that presented by Marconi \etal\.
For this cluster differential reddening may be important. Schlegel
\etal\ (\cite{schlegel98}) estimate $E_{B-V}=0.087$ mag.
Contamination of the MS by field stars is seen primarily as a vertical
sequence around $(B-V)\simeq0.6$. However, the group of stars bluer
than $(B-V)\simeq0.4$ and fainter than $V\sim21$ can conceivably be
the tip of the white dwarf cooling sequence associated to the cluster.
These stars are particularly interesting for follow-up observations
with FLAMES. At magnitudes brighter than the TO ($V=14.7$) a small
number of blue-straggler candidates are also seen. More candidates can
be found if one considers a larger area than the one shown in
Figure~\ref{F_cmd_all2}.

Several issues related to this cluster could be addressed with
FLAMES. In particular, a better and direct determination of its
metallicity.  Friel \& Janes (\cite{friel93}) measured
[Fe/H]$=-0.52\pm0.07$ from medium-resolution spectroscopy of five
cluster members, while Marconi
\etal\ using different stellar evolutionary models can fit the
observed data either with values of  metallicity consistent with
that determined spectroscopically or with solar values.

\void{ Second, to
investigate the claim of Carraro \& Chiosi (\cite{carraro94}) that
this cluster has not moved very far from its birthplace. which would
allow a better understanding of the effects of the environment in the
evolutionary history of these systems.}

\subsubsection{Small Magellanic Cloud fields}

The Magellanic Clouds are ideal targets for the study of galaxy
formation and evolution.  Their proximity has enabled ground-based
observations to unveil their oldest populations and resolve them into
individual stars.  Moreover, both the LMC and SMC contain young, rich
cluster systems that span a wide range of ages and metallicities,
evidence of their extended star formation history.  The young clusters
offer a unique opportunity to study the properties of massive stars
and to compare them with predictions of stellar evolutionary models.
Such studies might provide insight into the physical processes
responsible for their evolution, such as the treatment of mixing.

This PF release includes two adjacent fields in the SMC.  The combined
$V$ image produced by the stacking of the two DEEP exposures of one of
the SMC fields (SMC~5)   is shown in Figure~\ref{F_smc05}.   The image
shows   a  uniformly distributed population   of  SMC ``field stars'',
presumably the old population, and a considerable number of young star
clusters.  In SMC~5 the following clusters can be identified: NGC~346,
NGC~330, IC~1611, NGC~306, NGC~299, OGLE~109, OGLE~119, OGLE~99. 
 From
an inspection of  a similar image  for  SMC~6 one finds the  clusters:
OGLE~124, IC~1624, OGLE~129, K50, OGLE~134, K54. These images show the
great contribution of a wide-field imager to the study of such nearby
systems. The two fields  of SMC being  released provide a large survey
of OB associations which can be immediately  observed with FLAMES. One
of  the fields (SMC~5) includes the  brightest HII region (NGC~346) of
the SMC.    Massey et al.   (\cite{massey89})  confirm the presence of
$33$ O stars, $11$  of which are of type  O6.5 or earlier.  Six  stars
were found in the  mass range $40-80$~M$_\odot$.  They  also identified
few  red   supergiants of   considerably lower  mass   ($15$~M$_\odot$)
concentrated in a spatially distinct subgroup. From a close inspection
of the image shown Figure~\ref{F_smc05} the existence of this distinct
subgroup  is       confirmed     at     $\alpha=00^h59^m05.2^s$    and
$\delta=-72^\circ09^m15.4^s$.

The CMD of the two SMC fields show a number of populations with
different ages and metallicities, ideal targets for FLAMES.  Note that
the CMD of SMC~5 reaches fainter magnitudes compared to that of
SMC~6. This is because only one DEEP $V$ image is available for
SMC~6. The most prominent feature of both SMC CMDs is the MS centered
around $(B-V)\simeq-0.2$, as expected from the low reddening towards
the SMC ($E_{B-V}=0.037$, Schlegel et al.  \cite{schlegel98}). The
broadening of the MS at faint magnitudes ($V\lsim20$) can be due to
several effects such as crowding and increasing photometric errors.
For magnitudes $V\lsim17$ the broadening is mainly caused by blue
supergiants.  The MS terminates around $V\simeq14.3$ while the blue
supergiants extend up to $V\simeq12.5$.  At redder colors ($0.5\lsim
(B-V)\lsim2.0$) there is another tilted sequence which corresponds to
the red end of the blue loop (red supergiants). In between the the
blue and red-end of this loop, there is an almost vertical sequence
starting at $V\simeq19.5$, $(B-V)\simeq0.7$, which extends at least to
$V\simeq13.5$.  This population is the so-called ``vertical clump''
(Gallart et al. \cite{gallart98}) formed by young (few hundred Myr to
$1$ Gyr old), core $He$-burning and intermediate-mass stars.  Note,
however, that this sequence might suffer contamination by foreground
stars and supergiants along the loop. At $(B-V)\sim 1.6$ and $V\sim17$
one sees the tip of a broad red giant branch (RGB). The width of the
RGB at the tip is $\sim0.45$~mag.  This is probably due to the
combined effects of age and metallicity.  Finally, even though hard to
identify, given the resolution of the Figure~\ref{F_cmd_all2}, the red
extension of the horizontal branch is visible at $V\sim19.5$ in the
color interval $0.4\lsim(B-V)\lsim0.6$. However, one finds no evidence
of a blue horizontal branch.

The CMD shown in Figure~\ref{F_cmd_all2} reflects primarily the
properties of the SMC field star population. However, as mentioned
above the two SMC fields observed so far contain at least 14 stellar
systems, whose CMDs can also be extracted from the same data. Despite
possible contamination problems, such analysis is of great interest to
study the star formation history of both cluster and field
populations.

\subsection {Future Work}
\label{future}

As discussed above the data presented here provides not only the means
for producing target lists for FLAMES but also a wealth of information
about a variety of stellar systems. Furthermore, the data from the
survey can be combined to other data sets which can greatly enhance
the scientific value of the survey.

As  an example, for  all the observed fields  the final color catalogs
were associated   with those   being  produced by   the 2MASS  survey.
Combined EIS  and 2MASS data  will provide a  six passband multi-color
data   sets.    As   an    illustration     the  upper  panel       of
Figure~\ref{F_cmdbkv_2mass} shows  the optical/infrared  $V,(B-K)$ CMD
obtained for M~67.   Among  many  interesting  features, it  is  worth
noticing how the MS of  M~67 (Figure~\ref{F_cmd_all1}) is resolved into
two well-defined parallel sequences, much  more clearly than using the
optical alone.      The binary sequence is    clearly   visible in the
magnitude interval $14.5\lsim V\lsim18$.  Moreover, and in addition to
the well-known gap at  $V\sim13$, the diagram  shows clear evidence of
another  gap at $V\sim16.15$  and $(B-K)\sim3.35$.  The lower panel of
Figure~\ref{F_cmdbkv_2mass} shows the $K,(B-K)$  CMD  for SMC~5.    In
contrast to  the optical  CMD, the  optical/infrared provides a  clear
view  of the   lower temperature  stars,  while the  hot  MS and  blue
supergiants   essentially  disappear.  In this    diagram both the red
supergiant and the  red giant sequences  are clearly visible  and more
easily distinguishable. Also   note the asymptotic  giant branch (AGB)
stars  ($K\lsim12.5$) starting  from $(B-K)\gsim5.5$  and extending to
extremely red  colors ($(B-K)\sim11.2$).   Stars found in  this region
resemble those found   by Nikolaev \&   Weinberg (2000), studying  the
2MASS  diagram  of   the  LMC,    and  interpreted   as    Carbon-rich
Thermally-Pulsating  AGB and long period variables.   In the blue part
of the diagram  some MS and  blue supergiants  are still  present, and
seem to be  separated.  In addition, there  is a hint for a population
along the blue  loop  ($K\sim15.5$  and $0.5<(B-K)<2.5$).   Thus,  the
color baseline  provided by  the combination  of  optical and infrared
data, is a  powerful tool in  separating the different populations  of
composite systems.

Lastly, combining the optical and infrared data may also allow for the
spectral   classification  of   objects  by  matching  the photometric
measurements  against    template    spectra  (Hatziminaoglou    \etal
2001). This  may help  further disentangle  different  populations and
search for particular types of stars.  Results from such analysis will
be presented in a subsequent paper of this series.

\begin{figure}
\centering
\caption{The optical/infrared color-magnitude diagram obtained for
M~67 combining  EIS and 2MASS data.}
\label{F_cmdbkv_2mass}
\end{figure}

\section{Summary}
\label{S_summary}

This paper presents the first set of fully-calibrated images and
catalogs of the PF survey. The data presented here have been used to
evaluate the observing strategy, data reduction, catalog production
and preparation of target lists for feeding the FLAMES fiber
positioner. Particular emphasis has been given to the accuracy of the
astrometric calibration. The analysis presented in this paper has
shown the need for some improvement in the software, currently
underway, in order to efficiently cope with the large amount of data
already available. From a comparison of the results obtained using
different reference catalogs one finds that while the USNO~2.0
provides adequate astrometric solutions, with accuracy within the
requirements of FLAMES, future reductions should be done with the
recently released GSC~2.2 catalog. The latter proved to be far
superior with less systematics and smaller uncertainties.

The color catalogs produced have been used to derive color-magnitude
diagrams for all the fields considered.  The CMDs span $13$ magnitudes
in $V$ which ensures a better selection of both bright and faint
targets for FLAMES. A comparison of the CMDs with those of other
authors show good agreement over the magnitude range in
common. However, the PF data are far superior given their extended
magnitude and areal coverage. Furthermore, combining these data with
ongoing surveys in other wavelengths gives additional leverage for the
identification of different stellar populations, of key importance for
envisioned FLAMES programs.

All  the specific software developed to  deal with the  PF survey data
are currently being incorporated into the EIS survey system framework.
 This should allow the efficient reduction of all of the remaining data
gathered by this survey.   The  fully calibrated images  and  catalogs
presented  in this paper will  be  used in the  commissioning phase of
FLAMES and can be requested from the URL ``http://www.eso.org/eis''.


%

\begin{acknowledgements}
We thank L. Pasquini, for having promoted the Pre-FLAMES survey and
for his constant encouragement. We also thank A. Renzini for his
continuing support of the EIS project.  F. Comeron, V. Hill, and
F. Primas are thanked for their support in the preparation of the
original proposal.  We would also like to acknowledge F. Ferraro,
V. Hill, J.  Krautter, J.  Mermillod, B.  Noerdstrom, S.  Ortolani, G.
Piotto, F.  Primas, S.  Randich and E.  Tolstoy for their suggestions
in the initial phase of selection of the Pre-FLAMES targets. YM thanks
the Universit\`a di Padova, E.V. Held, G.Piotto \& S. Ortolani.  SZ,
thanks the Osservatorio di Trieste. We also thank M. Zoccali for
careful reading of the paper.

\end{acknowledgements}

\end{document}